\documentclass[twocolumn,aps,showpacs,preprintnumbers,lettersize]{revtex4}
\usepackage{amssymb}
\usepackage{graphicx}
\usepackage{dcolumn}
\usepackage{bm}

\begin{document}

\title{Pseudogap and spin fluctuations in the normal state of electron-doped
cuprates}
\author{B.~Kyung$^{1}$, V.~Hankevych$^{1,2}$, A.-M.~Dar\'{e}$^{3}$, and
A.-M.~S.~Tremblay$^{1}$}
\affiliation{$^{1}$D\'{e}partement de physique and Regroupement qu\'{e}b\'{e}cois sur les
mat\'{e}riaux de pointe, Universit\'{e} de Sherbrooke, Sherbrooke, Qu\'{e}%
bec J1K 2R1, Canada\\
$^2$Department of Physics, Ternopil State Technical University, 56 Rus'ka
St., UA-46001 Ternopil, Ukraine\\
$^{3}$L2MP, 49 rue Joliot Curie BP 146, Universit\'{e} de Provence, 13384
Marseille, Cedex 13, France}
\date{\today }

\begin{abstract}
We present reliable many-body calculations for the $t$-$t^{\prime }$-$%
t^{\prime \prime }$-$U$ Hubbard model that explain in detail the results of
recent angle-resolved photoemission experiments on electron-doped
high-temperature superconductors. The origin of the pseudogap is traced to
two-dimensional antiferromagnetic spin fluctuations whose calculated
temperature dependent correlation length also agrees with recent neutron
scattering measurements. We make specific predictions for photoemission, for
neutron scattering and for the phase diagram.
\end{abstract}

\pacs{74.72.-h, 71.10.Fd, 71.27.+a, 79.60.-i}
\maketitle



High-temperature superconductors (HTSC) still present one of the main
contemporary challenges to condensed matter physics. While single-particle
excitations, in the standard theory, are described by the quasiparticle
(Fermi liquid) concept, that approach fails in HTSC. Quasiparticles can
disappear altogether on certain segments of the would-be Fermi surface \cite%
{Damascelli:2003}. This in turn leads to anomalous properties in transport
and in thermodynamic data that are collectively referred to as pseudogap
phenomena. Pseudogap phenomena occur both in hole-doped (h-doped) cuprates
and in the electron-doped (e-doped) ones. Although the majority of research
has focused on h-doped materials, experimental activity on e-doped cuprates
has been steadily increasing in the last few years, as the community
realizes the importance of looking at the big picture for HTSC. Strong
electron-electron interactions and low dimensionality are the main stumbling
blocks for theories of HTSC. In this context, one can argue that it is the
lack of controlled theoretical approximations and consequent lack of
quantitative predictions and detailed agreement with experiment that are the
main reasons why there is no consensus on the correct theory in this field.
We will show in this paper that the somewhat weaker coupling in e-doped HTSC
leads to an unprecedented opportunity to obtain detailed agreement between
theory and experiment.

We use the single-band Hubbard model on a square lattice, which contains a
repulsive local interaction $U$ and a kinetic contribution of bandwidth $W$
that is fitted to band structure using nearest $t$, second-nearest $%
t^{\prime }$, and third-nearest $t^{\prime \prime }$ neighbor hoppings.
Reliable calculations for the Hubbard model can be done when the expansion
parameter $U/W$ becomes less than unity. As was pointed out by several
groups~\cite{Kusko:2002,Kyung:2002,Markiewicz:2003,Senechal:2003}, this is
precisely the situation that occurs in e-doped HTSC as doping is increased
towards optimal doping. Smaller values of $U/W$ physically come from better
screening in e-doped systems \cite{Senechal:2003}. Note that the pseudogap
mechanism discussed here can be different from that occuring at strong
coupling \cite{Senechal:2003}.

We calculate the momentum $\vec{k}$ and energy $\omega $ dependent
single-particle spectral weight $A(\vec{k}\mathbf{,}\omega )$ and find that
it is in detailed agreement with Angle Resolved Photo-Emission Spectroscopy
(ARPES)~on Nd$_{2-x}$Ce$_{x}$CuO$_{4}$ \cite%
{Armitage:2001,Armitage:2002,Damascelli:2003}. In particular, we explain the
pseudogap observed in e-doped cuprates at binding energies of the order of a
few hundred meV. To show that we correctly handle the physics of
antiferromagnetic (AFM) fluctuations that is behind that pseudogap, we
successfully compare our temperature-dependent correlation length to that
measured by neutron scattering~on Nd$_{2-x}$Ce$_{x}$CuO$_{4}$ \cite%
{Mang:2003,Matsuda:1992}. None of the recent theoretical works on e-doped
cuprates \cite%
{Kusko:2002,Kyung:2002,Markiewicz:2003,Senechal:2003,Tohyama:2001,Kusunose:2003}
have obtained momentum, frequency as well as \textit{temperature} dependent
properties that are in detailed agreement with both ARPES and neutron data.
We also find good estimates for the N\'{e}el temperature~\cite{Mang:2003}
and for the \textit{d}-wave superconducting transition temperature~\cite%
{Blumberg:2002}. A few of our predictions are that: (a) The ARPES pseudogap
disappears with increasing temperature when the AFM correlation length $\xi $
becomes smaller than the single-particle thermal de Broglie wavelength $\xi
_{th}$. The corresponding pseudogap temperature $T^{\ast }$ is close to
that~found in optical experiments \cite{Onose:2001}. (b) In the pseudogap
regime, the characteristic spin fluctuation energy in neutron scattering
experiments is smaller than the thermal energy (renormalized classical
regime) and the spin fluctuations are overdamped near the pseudogap
temperature. (c) There is no large energy ($\gtrsim $ $100\;$meV) pseudogap
at dopings larger than that of the quantum critical point (QCP), whose
location we also calculate and which coincides with that observed
experimentally for the antiferromagnetic to paramagnetic QCP \cite%
{Mang:2003,Greene:2004}.

Since we are in a non-perturbative regime, even with 
\mbox{$U/W\lesssim 1$}%
, calculations are still a challenge. We use the non-perturbative
Two-Particle Self-Consistent approach (TPSC) valid in that regime. The
reasons are as follows. TPSC satisfies at the same time the Pauli principle,
the Mermin-Wagner theorem and has a self-energy that is consistent with the
sum-rule relating potential energy to single-particle properties~\cite%
{vct94,vt97,malkpvt00,allen03}. By contrast with other approaches~\cite%
{Markiewicz:2003}, TPSC also automatically takes into account quantum
fluctuations that renormalize $U$ (Kanamori screening). Results for the
charge and spin structure factors, spin susceptibility and double occupancy
obtained with the TPSC scheme satisfy conservation laws. As a theory then,
TPSC fulfills crucial physical constraints. In addition, the above
quantities have been found to be in quantitative agreement with Quantum
Monte Carlo (QMC) and other numerical methods~\cite%
{vct94,vt97,malkpvt00,allen03,vdcvt95,klpt03} for both the nearest-neighbor~%
\cite{vct94,vt97} and next-nearest-neighbor~\cite{vdcvt95} Hubbard models in
two dimensions. In particular, the agreement with QMC for single-particle
properties has been verified~\cite{vt97,malkpvt00,klpt03} even as we enter
the pseudogap regime at low enough temperature. An internal accuracy check 
\cite{vt97} allows one to decide on the validity of the results in cases
where QMC or other exact results are not available as references. We do not
consider the case of very large correlation length, where the quantitative
details of TPSC become less accurate. Therefore, we are always within the
domain where the approach is reliable.


\begin{figure}[tbp]
\includegraphics[width=8.5cm]{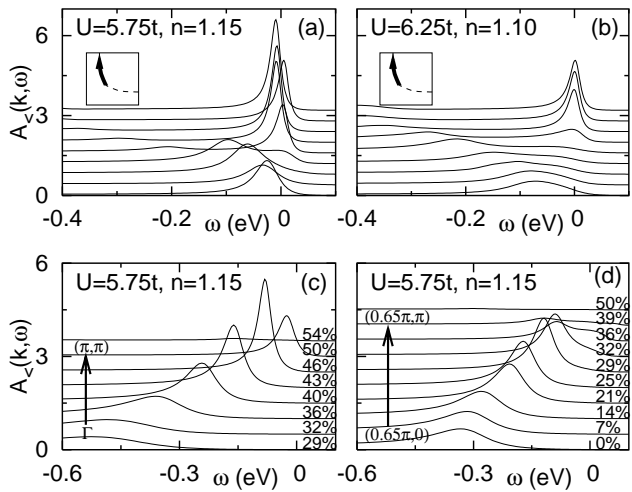}
\caption{Energy distribution curves $A_{<}(\vec{k},\protect\omega )\equiv A(%
\vec{k},\protect\omega )f(\protect\omega )$ along the Fermi surface shown in
the insets for (a) $n=1.15$, $U=5.75t$ (b) $n=1.10$, $U=6.25t$. Lines are
shifted by a constant for clarity. Spectra along two other directions for $%
n=1.15$ are also shown: (c) $(0,0)-(\protect\pi ,\protect\pi )$ and (d) $%
(0.65\protect\pi ,0)-(0.65\protect\pi ,\protect\pi )$. See also the
corresponding experimental data in Fig.~2 of Ref.~\protect\cite%
{Armitage:2002} and Fig.~1 of Ref.~\protect\cite{Armitage:2001}. Band
parameters are $t^{\prime }=-0.175t$, $t^{\prime \prime }=0.05t$, $t=350$
meV while $T=t/20$.}
\label{spectral_function_fig}
\end{figure}

\paragraph{Energy dispersion curves}

One of the most striking features of ARPES results in optimally e-doped
cuprates is the disappearance of $\omega =0$ excitations near the
intersection of the non-interacting Fermi surface with the AFM zone
boundary. These pseudogaped points in the Brillouin zone are called hot
spots. Band structure parameters that are input in the calculation lead to
intersection of the Fermi surface with the AFM zone boundary. Throughout the
paper, we present results for $t^{\prime }=-0.175t$ and $t^{\prime \prime
}=0.05t$, but very similar results are obtained for another choice of
parameters~\cite{Markiewicz:2003} $t^{\prime }=-0.275t$ and $t^{\prime
\prime }=0.0$~\cite{NoteAccuracy}. There are no phenomenological parameters,
such as mode coupling constants or additional renormalizations of $U$, as
can occur in other theories. For actual comparisons with experiment, we take 
$t=350$ meV~\cite{Kim:1998} since this allows us to find agreement with the
experimentally observed overall ARPES dispersion.

We present in Fig.~\ref{spectral_function_fig} the single-particle spectral
weight for two fillings, $n=1.15$ and $1.10$. Here, $A(\vec{k},\omega )$ is
multiplied by the Fermi-Dirac distribution function $f(\omega )$ to allow
direct comparison with experiment on Nd$_{2-x}$Ce$_{x}$CuO$_{4}$ $(n=1+x)$~%
\cite{Armitage:2001,Armitage:2002}. One should also recall that experimental
data contain a background at large binding energy whose origin is most
probably extrinsic~\cite{Kaminski:2003}. The results that we present in the
next two figures, Figs.~\ref{spectral_function_fig} and \ref%
{fermisurface_plot_fig}(a-b), are at temperature $T=t/20$ ($200$ K) that is
high compared with phase transitions~\cite{Mang:2003} but that is below the
pseudogap temperature $T^{\ast }$. We have verified that they remain
essentially unchanged~\cite{NoteTemperatureEffect} as $T$ is decreased
towards the range where experiments have been performed~\cite%
{Armitage:2001,Armitage:2002}.

Our results for $15\%$ doping (Fig.~\ref{spectral_function_fig}(a)) are in
remarkable agreement with ARPES data on reduced e-doped cuprates~\cite%
{Armitage:2001,Armitage:2002} when we use $U=5.75t$. In ARPES,
superconductivity leads only to a $2$ meV shift of the leading edge~\cite%
{Armitage:2001} so it would not be observable on the scale of Fig.~\ref%
{spectral_function_fig}. We see that $A_{<}(\vec{k}_{F},\omega )$ is peaked
near $(\pi ,0)$ and $(\pi /2,\pi /2)$ and is pseudogaped at hot spots.
Smaller values of $U$ can be below the critical $U_{c}$ necessary for spin
fluctuations to build up in the presence of frustrating $t^{\prime }$ and $%
t^{\prime \prime }.$ At $U=5.25t$, for instance, sharp quasiparticle peaks
appear everywhere along the Fermi surface (not shown here). For larger $U,$
the spectral weight $A_{<}(\vec{k}_{F},\omega )$ is strongly suppressed near 
$(\pi /2,\pi /2)$~because scattering by spin fluctuations is stronger and
extends over a broader region in $\mathbf{k}$ space \cite{Kyung:2002}. For $%
10\%$ e-doped cuprates, on the other hand, the best fit to experiments is
found for $U=6.25t$ (Fig.~\ref{spectral_function_fig}(b)), not for $U=5.75t$%
. At $U=6.25t$ the weight $A_{<}(\vec{k}_{F},\omega )$ near $(\pi /2,\pi /2)$
is shifted away from the Fermi energy (pseudogap), in good agreement with
the experimental results~\cite{Armitage:2002}, while at $U=5.75t$ (not shown
here), a pseudogap would not appear near $(\pi /2,\pi /2)$ \cite%
{NoteAlternateFit}. An increase of $U$ with decreasing doping is also
expected from the fact that $U=6.25t$ is not large enough to produce the
observed \cite{Uchida:1989} Mott gap at half-filling. Our calculated $A_{<}(%
\vec{k},\omega )$ along some selected directions also agrees well with ARPES
data~\cite{Armitage:2001}. Along $(0,0)-(\pi ,\pi )$ (Fig.~\ref%
{spectral_function_fig}(c)) a sharp peak approaches the Fermi energy near $%
\vec{k}_{F}$, while along $(0.65\pi ,0)-(0.65\pi ,\pi )$ (Fig.~\ref%
{spectral_function_fig}(d)) the peak does not go through the Fermi energy
and seems to bounce back near $\vec{k}_{F}$. The peak position where
bouncing occurs may be used to obtain a rough estimate of pseudogap size $%
\Delta _{PG}$. For $15\%$ and $10\%$ dopings we find about $0.1$ and $0.3$
eV, respectively.

\begin{figure}[tbp]
\includegraphics[width=7.0cm]{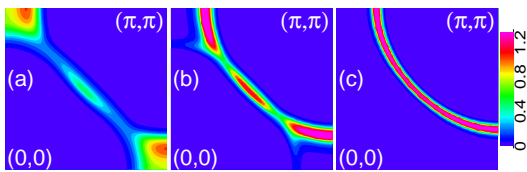}
\caption{(color) Fermi surface plot for (a) $U=6.25t$, $n=1.10,$ $T=t/20$
(b) $U=5.75t$, $n=1.15,$ $T=t/20$ (c) $U=5.75t$, $n=1.20,$ $T=t/50$ in the
first quadrant of the Brillouin zone. See the corresponding experimental
plots in Fig.~3 of Ref.~\protect\cite{Armitage:2002}.}
\label{fermisurface_plot_fig}
\end{figure}

Calculated Fermi surface plots $A_{<}(\vec{k},0)$ are obtained by
integrating $A(\vec{k},\omega )f(\omega )$ in a window $(-0.2t,0.1t)$,
comparable to that of experiment~\cite{Armitage:2002}. The overall features
are not sensitive to this choice for the parameters used here. For $10\%$
doping (Fig.~\ref{fermisurface_plot_fig}(a)), strong AFM fluctuations cause
the suppression of spectral weight not only at hot spots, but also along a
large segment of the Fermi surface near $(\pi /2,\pi /2)$. As a consequence,
at $10\%$ doping the Fermi surface plot looks as if it was composed of
pockets near $(\pi ,0)$. As shown in Fig.~\ref{fermisurface_plot_fig}(b),
for $15\%$ doping the spectral weight at $\omega =0$ is strongly suppressed
by AFM fluctuations only at hot spots. All these results are observed in
ARPES data~\cite{Armitage:2001,Armitage:2002} for e-doped cuprates. These
features follow from $\vec{k}$ dependent scattering rates due to AFM
fluctuations.

\paragraph{Pseudogap temperature and quantum critical point}

Within TPSC, it has been predicted analytically that the pseudogap in the
spectral function occurs at $T^{\ast }$ when points of the two-dimensional
Fermi surface can be connected by the AFM wave vector~\cite{v97} and when $%
\xi $ begins to exceed $\xi _{th}=\hslash v_{F}/\pi k_{B}T$ at these points~%
\cite{v97,vt97,malkpvt00}. The pseudogap is a precursor of the long-range
AFM state in two dimensions. Decreasing $T$ below $T^{\ast }$ increases $\xi 
$ beyond $\xi _{th}$ and does not produce large changes in the spectra \cite%
{NoteTemperatureEffect}. In our calculations, the AFM-induced $T^{\ast }$
ends at the same value of doping as the QCP where $T=0$ two-dimensional
antiferromagnetism disappears. For $x$ below the QCP, the $T=0$ magnetic
order can be commensurate or incommensurate although at $T^{\ast }$ the
fluctuations are commensurate. The two-dimensional QCP, which may be masked
by a superconducting Kosterlitz-Thouless transition \cite{Note_KT} or by
three-dimensional ordered phases, is near $18\%$, much smaller than the
naive percolation threshold~\cite{Mang:2003}. At the QCP we find $\omega /T$
scaling in the self-energy and $\xi \sim T^{-1/z}$ with dynamical exponent $%
z=2$. This asymptotically implies $\xi <\xi _{th}$ and hence no pseudogap at
the QCP. Fig.~\ref{fermisurface_plot_fig}(c) illustrates our prediction that
hot spots in $A_{<}(\vec{k},0)$ disappear beyond the QCP even at very low
temperature. We are also predicting that the pseudogap observed in ARPES
energy distribution curves at low temperature should persist almost
unchanged up to the $T^{\ast }$ indicated by the filled circles in Fig.~\ref%
{phasediagram_fig}. As temperature increases above $T^{\ast }$, the
pseudogap in energy distribution curves begins to disappear and red segments
in Fig. \ref{fermisurface_plot_fig} will enlarge gradually until one
recovers a continuous Fermi surface at around $2T^{\ast }$.

\begin{figure}[tbp]
\includegraphics[width=7.0cm]{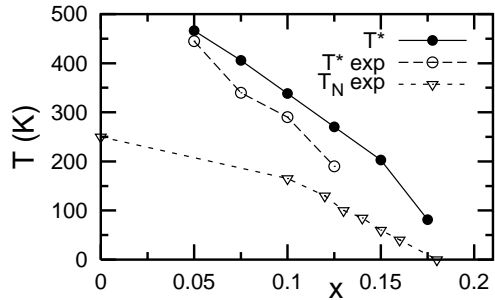}
\caption{Pseudogap temperature $T^{\ast }$ (filled circles denote calculated 
$T^{\ast }$, empty ones experimental data extracted from optical
conductivity~\protect\cite{Onose:2001}). Empty triangles are experimental N%
\'{e}el temperatures $T_{N}$. The samples are reduced~\protect\cite%
{Mang:2003}.}
\label{phasediagram_fig}
\end{figure}

The empty circles in Fig.~\ref{phasediagram_fig} are the observed pseudogap
temperatures $T^{\ast }$ extracted from optical conductivity data by Onose 
\emph{et al.}~\cite{Onose:2001}. This $T^{\ast }$ follows closely the $%
T^{\ast }$ that we predict for ARPES and scales with the observed $T_{N}$
curve (approximately two times larger) supporting the AFM fluctuation origin
of the pseudogap in the normal state of e-doped cuprates. Onose \emph{et al.}%
~\cite{Onose:2001} found that $\Delta _{PG}$ and $T^{\ast }$ satisfy the
relation $\Delta _{PG}\approx 10k_{B}T^{\ast }$. Our estimation of the ratio
of $\Delta _{PG}$ to $T^{\ast }$ leads to $10$ and $7.5$ for $10\%$ and $%
15\% $ dopings, respectively, in rough agreement with the above experiment.
Note that $T^{\ast }$ is considerably decreased compared with the mean-field 
$T_{N}$ by both Kanomori screening and thermal fluctuations, two effects
taken into account by TPSC.

\paragraph{Antiferromagnetic correlation length}

\begin{figure}[tbp]
\includegraphics[width=7.0cm]{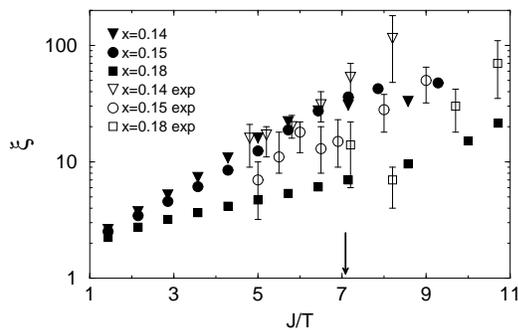}
\caption{Semi-log plot of the AFM correlation length (in units of the
lattice constant) against inverse temperature (in units of $J=125$ meV) for $%
x=$0.14, 0.15, and 0.18. Filled symbols denote calculated results and empty
ones experimental data~of Refs.\protect\cite{Mang:2003} and \protect\cite%
{Matsuda:1992} ($x=0.15$). The arrow indicates $T=t/20$ where most spectral
quantities in Figs.~\protect\ref{spectral_function_fig} and~\protect\ref%
{fermisurface_plot_fig} are calculated.}
\label{corrlength_fig}
\end{figure}

To demonstrate convincingly that our picture is consistent, we need to show
that our results also agree with the experimentally observed AFM $\xi $. Mang%
\textit{\ et al.}~\cite{Mang:2003} and Matsuda \textit{et al. }\cite%
{Matsuda:1992} have obtained neutron scattering results for $\xi (T)$ at
various dopings for oxygenated samples of Nd$_{2-x}$Ce$_{x}$CuO$_{4}$. They
find an exponential dependence of $\xi (T)$ on inverse temperature. This is
illustrated on an Arrhenius plot in Fig.~\ref{corrlength_fig} along with our
calculated $\xi (T)$. Analytical results for TPSC~\cite{vct94} indeed show
that the pseudogap occurs in the renormalized classical regime where the
characteristic spin fluctuation energy is less than the thermal energy and
where $\xi (T)\sim \exp (C/T)$ with $C$ being generally a weakly temperature
dependent constant. The temperature at which $A_{<}(\vec{k},\omega )$ has
been calculated, $T=t/20$ denoted by an arrow in Fig.~\ref{corrlength_fig},
is still in the exponential regime. The value of $\xi (T)$ at that
temperature for optimal doping is $20$ to $30$ lattice constants (long but
finite). For $10\%$ doping (not shown here), the calculated $\xi (T)$ is
also in good agreement at high temperatures but the agreement becomes less
satisfactory at low temperatures since the quantitative details of TPSC
become less accurate for very large $\xi $. Also, one should allow for some
small uncertainty in the doping~\cite{Mang:2003} or in the parameters of the
Hubbard model because neutron experiments were performed on oxygenated (as
grown) samples while ARPES was done on reduced ones~\cite{Note-Oxygen}. We
predict that dynamic neutron scattering will find overdamped spin
fluctuations as we enter the pseudogap regime and that the characteristic
spin fluctuation energy will be smaller than $k_{B}T$ whenever a pseudogap
is present. Equality will occur well above $T^{\ast }$.

\paragraph{Long-range order}

In the extended TPSC approach~\cite{Kyung:2002} AFM fluctuations can both
help and hinder $d$-wave superconductivity, depending on the strength of the
pseudogap. For the case at hand, a finite $T_{c}$ appears below $18\%$
doping, and increases with decreasing doping. We cannot study at this point
the effect of the appearance of long-range N\'{e}el order on
superconductivity but the estimated value of $T_{c}$ near optimal doping is
roughly the experimental result~\cite{Armitage:2001,Blumberg:2002}. By
adding hopping $t_{z}=0.03t$ in the third direction, we also find a N\'{e}el
temperature that is in fair agreement with the experimental data by Mang 
\textit{et al.}~\cite{Mang:2003} for $x\geq 0.14$.

In contrast to e-doped cuprates, in h-doped systems a pseudogap opens up
near $(\pi ,0)$ while spectral weight persists near $(\pi /2,\pi /2)$. TPSC
can also qualitatively reproduce this result but the AFM correlation length
in this case disagrees with experiment. A different short range mechanism
can come into play in the strong coupling limit more appropriate for h-doped
cuprates \cite{Senechal:2003}.

In conclusion, we have shown that ARPES~spectra \cite%
{Armitage:2001,Armitage:2002} and the antiferromagnetic correlation length
obtained by neutron scattering~\cite{Mang:2003,Matsuda:1992} in e-doped
cuprates can all be theoretically explained in detail within the Hubbard
model in the weak to intermediate coupling regime. We have made several
predictions for future experiments. The physical picture of the normal state
that emerges is that of electrons on a planar lattice scattering off
quasi-static antiferromagnetic fluctuations. These have large phase space in
two dimensions so that quasiparticles in regions of the Fermi surface that
can be connected by the antiferromagnetic wave vector do not survive
scattering by these low-energy bosons~\cite{Vilk:1995,vt97,v97}. The
electron-doped cuprates appear to be the first high-temperature
superconductors whose normal state lends itself to such a detailed
theoretical description.

We acknowledge useful discussions with N.P.~Armitage, P.~Fournier,
M.~Greven, R.S. Markiewicz, Y.~Onose, M. Rice, D.~S\'{e}n\'{e}chal and L.
Taillefer. The present work was supported by NSERC Canada, FQRNT Qu\'{e}bec,
CIAR and the Tier I Canada Research Chair Program (A.-M.S.T.). A.-M.S.T.
acknowledges the hospitality of Universit\'{e} de Provence and of Yale
University where part of this work was performed and, in the latter case,
supported under NSF Grant-0342157.


\begin{thebibliography}{99}
\bibitem{NoteAlternateFit} One can obtain nearly the same results at $n=1.1$
with the same value $U=5.75t$ as that used for $n=1.15$ if the band
parameters are changed, for example, to $t^{\prime }=-0.13t$ and $t^{\prime
\prime }=0.06t$. This change of order $20\%$ does not appear realistic.
\bibitem{Note_KT} When this occurs, additional superconducting precursor
effects may contribute to the pseudogap.
\end{thebibliography}
\end{document}